
\newif\iffigs\figstrue

%
\let\useblackboard=\iftrue
%
%
\newfam\black

\input harvmac.tex

\iffigs
  \input epsf
\else
  \message{No figures will be included.  See TeX file for more
information.}
\fi

\def\Title#1#2{\rightline{#1}
\ifx\answ\bigans\nopagenumbers\pageno0\vskip1in%
\baselineskip 15pt plus 1pt minus 1pt
\else
\def\listrefs{\footatend\vskip 1in\immediate\closeout\rfile\writestoppt
\baselineskip=14pt\centerline{{\bf References}}\bigskip{\frenchspacing%
\parindent=20pt\escapechar=` \input
refs.tmp\vfill\eject}\nonfrenchspacing}
\pageno1\vskip.8in\fi \centerline{\titlefont #2}\vskip .5in}

\ifx\answ\bigans\def\tcbreak#1{}\else\def\tcbreak#1{\cr&{#1}}\fi
\useblackboard
\message{If you do not have msbm (blackboard bold) fonts,}
\message{change the option at the top of the tex file.}
\font\blackboard=msbm10 
\font\blackboards=msbm7
\font\blackboardss=msbm5
\textfont\black=\blackboard
\scriptfont\black=\blackboards
\scriptscriptfont\black=\blackboardss
\def\Bbb#1{{\fam\black\relax#1}}
\else
\def\Bbb#1{{\bf #1}}
\fi
%
\def\yboxit#1#2{\vbox{\hrule height #1 \hbox{\vrule width #1
\vbox{#2}\vrule width #1 }\hrule height #1 }}
\def\fillbox#1{\hbox to #1{\vbox to #1{\vfil}\hfil}}
\def\ybox{{\lower 1.3pt \yboxit{0.4pt}{\fillbox{8pt}}\hskip-0.2pt}}
\def\np#1#2#3{Nucl. Phys. {\bf B#1} (#2) #3}
\def\npfs#1#2#3#4{Nucl. Phys. {\bf B#1} [FS #2] (#3) #4}
\def\pl#1#2#3{Phys. Lett. {\bf #1B} (#2) #3}

\def\mpl#1#2#3{Mod. Phys. Lett. {\bf #1} (#2) #3}

\def\comments#1{}

\def\half{{1\over 2}}

\def\Im{{\rm Im\hskip0.1em}}

\def\ket#1{|#1\rangle}

\def\a{\alpha}

\def\II{\relax{I\kern-.07em I}}

\def\IZ{\relax\ifmmode\mathchoice
{\hbox{\cmss Z\kern-.4em Z}}{\hbox{\cmss Z\kern-.4em Z}}
{\lower.9pt\hbox{\cmsss Z\kern-.4em Z}}
{\lower1.2pt\hbox{\cmsss Z\kern-.4em Z}}\else{\cmss Z\kern-.4em
Z}\fi}
\def\IB{\relax{\rm I\kern-.18em B}}

\def\ID{\relax{\rm I\kern-.18em D}}
\def\IE{\relax{\rm I\kern-.18em E}}
\def\IF{\relax{\rm I\kern-.18em F}}
\def\IG{\relax\hbox{$\inbar\kern-.3em{\rm G}$}}
\def\IGa{\relax\hbox{${\rm I}\kern-.18em\Gamma$}}
\def\IH{\relax{\rm I\kern-.18em H}}
\def\II{\relax{\rm I\kern-.18em I}}
\def\IK{\relax{\rm I\kern-.18em K}}
\def\IP{\relax{\rm I\kern-.18em P}}

\useblackboard
\def\IZ{\relax\Bbb{Z}}
\fi

\font\cmss=cmss10 \font\cmsss=cmss10 at 7pt
\def\IR{\relax{\rm I\kern-.18em R}}

\def\Im{{\rm Im\ }}
\def\BR{\IR}
\def\BZ{\IZ}
\def\BR{\IR}



\def\lim{{lim}}

\input epsf

\def\SUSY#1{{{\cal N}= {#1}}}                   

\def\wdg{{\wedge}}                              


\def\inv#1{{1\over{#1}}}                              


\def\MS#1{{{\bf S}^{#1}}}               
\def\MT#1{{{\bf T}^{#1}}}               

\def\MHT#1{{{\hat{\bf T}}^{#1}}}               



\def\hepth#1{{\tt hep-th/{#1}}}

\def\Id{{\bf I}}                             

\def\com#1#2{{\lbrack {#1},{#2} \rbrack}}      



\def\trp#1{{{\rm tr}\{ {#1} \} }}            

\def\u{{\mu}}
\def\v{{\nu}}

\def\lam{{\lambda}}




\Title{\vbox{\baselineskip12pt\hbox{hep-th/9611202, PUPT-1668}}}
{\vbox{
\centerline{Branes, Fluxes and Duality in M(atrix)-Theory}}}
\centerline{Ori J. Ganor, Sanjaye Ramgoolam and
            Washington Taylor IV}
\smallskip
\smallskip
\centerline{Department of Physics, Jadwin Hall}
\centerline{Princeton University}
\centerline{Princeton, NJ 08544, USA}
\centerline{\tt origa,ramgoola,wati@puhep1.princeton.edu}
\bigskip
\bigskip
\bigskip
\noindent
We use the T-duality transformation which relates M-theory
on $T^3$ to M-theory on a second $T^3$ with inverse volume
to test the Banks-Fischler-Shenker-Susskind suggestion for
the matrix model description of M-theory.
We find evidence that T-duality is realized as S-duality
for $U(\infty)$ $N=4$ Super-Yang-Mills in 3+1D. We argue that
Kaluza Klein states of gravitons correspond to electric fluxes,
wrapped membranes become magnetic fluxes and instantonic membranes
are related to Yang-Mills instantons.
The T-duality transformation of gravitons
into wrapped membranes is interpreted as the duality between
electric and magnetic fluxes.
The identification of M-theory T-duality as SYM S-duality provides
a natural framework for studying the M-theory 5-brane as the
S-dual object to the unwrapped membrane.
Using the equivalence between compactified M(atrix) theory and SYM, we
find a natural candidate for a description of the
light-cone 5-brane of M-theory directly in terms of matrix variables,
analogous to the known description of the M(atrix) theory membrane.

\Date{November, 1996}

\lref\BFSS{T. Banks, W. Fischler, S. H. Shenker and L. Susskind,
  {\it M Theory As A Matrix Model: A Conjecture,} \hepth{9610043}}

\lref\hultow{C. Hull and P. K. Townsend, \np{438}{1995}{109}, \hepth{9410167}}

\lref\witvar{E. Witten,
  \np{443}{1995}{85}, \hepth{9503124}}

\lref\WitBST{E. Witten,
{\it Bound States of Strings and $p$-Branes},  
\hepth{9510135}}

\lref\NDW{B. de Wit, J. Hoppe and H. Nicolai,
  \npfs{305}{23}{1988}{545};
  B. de Wit, M. Luscher and H. Nicolai, \np{320}{1989}{135}}

\lref\wati{W. Taylor,
  {\it D-brane field theory on compact spaces,} \hepth{9611042}}

\lref\Sus{L. Susskind,
  {\it T Duality in M(atrix) Theory and S Duality in Field Theory,}
  \hepth{9611164}}

\lref\Sen{A. Sen,
  \mpl{11}{1996}{827}, \hepth{9512203}}

\lref\BHO{E. Bergshoeff ,C. M. Hull ,T. Ortin,
  \np{452}{1995}{547}, \hepth{9504081}}

\lref\vipul{V. Periwal,
  {\it Matrices on a point as the theory of everything,}
  \hepth{9611103}}

\lref\Ah{O. Aharony,
  {\it String theory dualities from M-theory,} \hepth{9604103}}

\lref\thft{G. `t Hooft,
\np{153}{1979}{141}}

\lref\BerDou{M. Berkooz and M. R. Douglas,
  {\it Five-branes in M(atrix) Theory,} \hepth{9610236}}

\lref\town{P. K. Townsend, 
\pl{373}{1996}{68}.  }

\lref\senmarg{A. Sen, {\it A Note on Marginally Stable Bound
   States in Type II String Theory,} 
\hepth{9510229}}

\lref\sch{J. H. Schwarz, 
\pl{B360}{1995}{13}.  }

\lref\asy{O. Aharony, J. Sonnenschein and S. Yankielowicz,
  \np{474}{96}{309}, \hepth{9603009}}

\lref\strom{A. Strominger, {\it Open p-Branes,} \hepth{9512059}}

\lref\christof{C. Schmidhuber,
\np{467}{96}{146}, \hepth{9601003}}

\lref\witteni{E. Witten,
\np{460}{1996}{541}, \hepth{9511030}.}

\lref\douglas{M. Douglas,
{\it  Branes within branes},
\hepth{9512077}; {\it Gauge fields and D-branes}, \hepth{9604198}.}

\lref\sg{G. Lifschytz and S. D. Mathur,
{\it Supersymmetry and Membrane Interactions in M(atrix) Theory},
\hepth{9612087}
}



\newsec{Introduction}

During the past two years, evidence has
been accumulating which indicates that a consistent
quantum theory (M-theory) underlies 11D supergravity \refs{\hultow,\witvar}.
Recently, an exciting conjecture for a microscopic
description of M-theory has been put forward
by Banks, Fischler, Shenker and Susskind \BFSS.
The BFSS model incorporates in a natural way the non-commutative
nature of microscopic space time \WitBST\ and the 
quantization of the membrane \NDW.

The authors of \BFSS\ have shown that their model contains many of the
features of M-theory: the supermembrane, correct graviton scattering
amplitudes, toroidal compactification and partial 11D Lorentz
invariance.  Further evidence for the BFSS conjecture was supplied in
\BerDou\ where the behavior of a membrane in a 5-brane background was
studied.  Questions which remain open include a general description of
compactification, an intrinsic description of a 5-brane and a complete
proof of 11D Lorentz invariance (a suggestion in this direction has
been made in \vipul).

The purpose of the present paper is to pass M(atrix)-theory through
one more test by considering its behavior under T-duality.  T-duality
relates compactified type IIA to compactified type IIB so in order to
get an ``automorphism'' of M-theory we need to compactify the type IIA
theory on $\MT{2}$ and apply T-duality twice.  This gives a connection
between M-theory on $\MT{3}$ with volume $V$ and M-theory on $\MT{3}$
with volume $1/V$. Under this duality, wrapped membrane states are
exchanged with Kaluza-Klein states of the graviton and unwrapped
membranes become wrapped 5-branes \refs{\Sen,\Ah}.

Our discussion begins with a review of the M(atrix)-theory description
of toroidal compactification given in \refs{\BFSS,\wati}.  The
resulting model is a large $N$ limit of $U(N)$ Super-Yang-Mills
theory.  We recast the derivation in a slightly different language and
explain how twisted sectors of the $U(N)$ bundle appear.  We then
argue that T-duality is realized as S-duality in the SYM theory and
that graviton $\Leftrightarrow$ membrane duality is related to
electric $\Leftrightarrow$ magnetic duality.  Finally, we discuss some
issues related to finding an explicit construction of the 5-brane in
M(atrix) theory.

The paper is organized as follows: Section 2 is a review of toroidal
compactification of the M(atrix)-model.  In Section 3 we relate
S-duality of $\SUSY{4}$ SYM to T-duality.  In Section 4, the
transformations of gravitons into membranes in M-theory are discussed.
We relate membranes to magnetic fluxes and gravitons with KK momentum
to electric fluxes.  In Section 5 we use the interpretation of the
membranes in terms of magnetic flux to obtain the energy and counting
of membrane states.  We also relate the Yang-Mills instanton to a
Euclidean membrane.  In Section 6 we suggest a formulation of a
5-brane wrapped around the light-cone directions in terms of BFSS
matrix variables satisfying a certain relation.  We also discuss the
implications of the T-duality/S-duality equivalence for the search for
a 5-brane in M(atrix) theory which extends along 5 transverse directions.


\newsec{Review of Compactification}

One way of understanding toroidal compactification of M(atrix)-theory
is by considering a sector of the $N\rightarrow\infty$
0-brane theory in which the $X$ matrices satisfy certain symmetry
conditions.
This description for compactification on the $d$-dimensional torus $\MT{d}$
can be given as follows \refs{\BFSS,\wati}: 
Infinite unitary matrices
$U_i$ are chosen for $i=1\dots d$ that commute with each other,
\eqn\uiuj{
U_i U_j = U_j U_i,
}
and generate a subgroup of $U(\infty)$ isomorphic to $\BZ^d$.
The compactified theory is given by restricting to
the subspace of $X$'s which are invariant under the $\BZ^d$ action:
\eqn\zdact{
U_i: X^\mu \longrightarrow U_i^{-1} X^\mu U_i + {{\bf e}_i}^\mu
}
where the ${\bf e}_i$'s form a basis of the lattice whose unit cell is
$\MT{d}$. 
The countably infinite dimensional vector space
on which the $X$'s act can be written as a tensor product 
\eqn\vvv{
V = V_N \otimes H^d
}
where $V_N$ is an $N$-dimensional space and $H$ is countably infinite
dimensional.
One can then take 
\eqn\mi{
U_i = \Id\otimes\Id_1\otimes\cdots \otimes \Id_{i-1}
\otimes S_i\otimes\Id_{i+1}\otimes\cdots\otimes\Id_d
}
where $\Id_j$ is the identity on the $j$th $H$
and $S_i$ is a shift operator
\eqn\mj{
(S_i)_{k,l} = \delta_{k+ 1,l}
}
(the indices  $k, l$ run over all integers $\BZ$).
By restricting the Lagrangian to the subspace of $X$'s invariant
under \zdact\ one obtains the Lagrangian for $(d+1)$ dimensional SYM theory.
The gauge fields are
\eqn\auf{
A^\u (\sum_{i=1}^d x_i {\bf \hat{e}}_i) =
\sum_{l_1,\dots,l_d\in\BZ} e^{2\pi i\sum l_i x_i} 
   X^\u_{(0,l_1)\cdots (0,l_d)}
}
where ${\bf \hat{e}}_i$ form a basis of the dual torus $\MHT{d}$.
When $d = 3$,
the inverse squared coupling constant is the volume of $\MT{3}$ (in
11-dimensional Planck units)
and the $(3+1)$-dimensional
theory is defined on the dual torus $\MHT{3}$.

For later use, we will recast the derivation in a different language.
We note that the resulting matrix $X^\u$ invariant under \zdact\
is just the matrix of the operator
\eqn\nabdef{
\nabla^\u = i\partial^\u  +A^\u
}
acting on fields in the fundamental representation
\eqn\mk{
\phi_k(\sum_{i=1}^d x_i {\bf \hat{e}}_i),\qquad k=1\dots N
}
which are sections of the trivial bundle over $\MHT{d}$.
The requisite form of the $X$ matrices
is obtained by writing $\nabla^\u$ in the Fourier basis of
$\hat{\phi}_k(n_1\dots n_d)$:
\eqn\ml{
\phi_k(\sum x_i {\bf \hat{e}}_i) =
\sum_{\{n_i\}} \hat{\phi}_{k}(n_1\dots n_d) e^{2\pi i \sum n_i x_i}.
}
The operators $U_j$ can be taken to act on sections by
\eqn\namea{
U_j \phi_k(\sum x_i {\bf \hat{e}}_i) = 
e^{2\pi i x_j} \phi_k(\sum x_i {\bf \hat{e}}_i)
}
{}From \nabdef\ it is clear that if we also take $X^\u$ for $\u=d+1,\dots ,9$
to be the matrices of the operators
\eqn\nameb{
\phi \longrightarrow \Phi^\u\phi
}
where $\Phi^\u$ are the scalar fields of SYM, then
the BFSS Lagrangian will reduce to the SYM Lagrangian.


\newsec{Checking T-duality}

A non-trivial duality of M-theory is obtained by compactifying
on $\MT{3}$, regarding the theory as type IIA on $\MT{2}$ and T-dualizing
twice (once along each direction of the $\MT{2}$).


\subsec{Review of T-duality for M-theory}

We will be using  eleven dimensional Planck units $l_p = 1$.
T-duality on M-theory is obtained from the relations (here 
$G_{\u\v}$ is the metric in coordinates $0,\dots,6,11$. Note
that our space-time coordinates are $x_0\dots x_9,x_{11}$):
\eqn\miia{
{{\rm M-theory} \over {G_{\u\v}; R_7,R_{8},R_{9}}}
=
{{\rm Type\ IIA} \over {R_{9} G_{\u\v}; R_{9}^{1/2} R_7, R_{9}^{1/2} R_{8};
\lam_{st} = R_{9}^{3/2}}}
}
together with type IIA T-duality twice (here $l_i$ are lengths in
string units):
\eqn\tdl{
{{\rm Type\ IIA} \over {g_{\u\v}; l_8, l_{9}; \lam}}
=
{{\rm Type\ IIA} \over {g_{\u\v}; l_8^{-1}, l_{9}^{-1}; 
    l_8^{-1} l_{9}^{-1}\lam}}
}
To obtain
\eqn\tdl{
{{\rm M-theory} \over {G_{\u\v}; R_7,R_{8},R_{9}}}
=
{{\rm M-theory} \over {V^{2/3}G_{\u\v}; 
    V^{-2/3}R_7, V^{-2/3}R_{8}, V^{-2/3}R_{9}}}
}
where $V=R_7 R_{8} R_{9}$ (the RHS actually comes out with 
$R_7$ and $R_8$ switched but in \tdl\ we have combined a 
reflection).
This can be generalized to slanted
$\MT{3}$'s and to include the 3-form $C_{\u\v\rho}$ \Sen:
\eqn\vv{
\tau = i V_{7,8,9} + C_{7,8,9} \longrightarrow -{1\over\tau}.
}

T-duality also acts non-trivially on D-brane states.
Let us denote the internal torus in directions $7,8,9$ by $\MT{3}$.
A state containing a graviton with momentum 
\eqn\namec{
\vec{p} = n_7 {\bf \hat{e}}_7 + n_8 {\bf \hat{e}}_8
         + n_9 {\bf \hat{e}}_9
}
(where ${\bf \hat{e}}_7, {\bf \hat{e}}_8, {\bf \hat{e}}_9$
are basis vectors for $\MHT{3}$ and we will assume $(n_7,n_8,n_9)$
to be relatively prime) is transformed by T-duality to a state 
with a membrane. When the T-duality 
is combined with   the $(78)$ reflection, 
the membrane is wrapped on the plane orthogonal to
$\vec{p}$ in the original $\MT{3}$.
This is easily seen by regarding the theory as type IIA on $\MT{2}$
and using the T-duality transformations of Dirichlet $p$-branes
into Dirichlet $(p\pm 2)$-branes together with the identification
of a D2-brane of type IIA with a membrane of M-theory and a 0-brane
of type IIA with a KK state of a graviton.

Next, let us take a membrane wrapped on, say, the 8-9 directions
and with momentum along the 7th direction.
Regarding it as an elementary string of type IIA wrapped on the 8th
direction and with momentum along the 7th direction we see that
after T-duality and the $(78)$
reflection, 
it becomes again a string wrapped on the 8th direction and with
momentum along the 7th direction -- so in M-theory it is another membrane
wrapped on the 8-9 directions and with momentum along the 7th direction.

In general, a membrane wrapped on a plane
orthogonal to $\vec{p}$ and with momentum 
\eqn\named{
\vec{q} = m_7 {\bf \hat{e}}_7 + m_8 {\bf \hat{e}}_8
         + m_9 {\bf \hat{e}}_9
}
($m_7,m_8,m_9$ are integers, not all zero)
becomes a membrane wrapped on a plane orthogonal to $\vec{q}$
(wrapped $gcd(m_7,m_8,m_9)$  times) and with momentum $\vec{p}$ \Sen.

We note that all these formulae are symmetric with respect to
the 3 directions of the $\MT{3}$ -- as they should be.
This symmetry is manifested by the fact that both membrane winding numbers
and Kaluza-Klein momenta are parameterized by a vector in the dual
lattice (whose unit cell is $\MHT{3}$). T-duality, composed with the 
reflection,  exchanges winding
with KK momentum.

Finally, we can take a membrane that is extended
in two non-compact directions, say $5,6$.
Regarding it as the D2-brane of type IIA, we see that it becomes
the D4-brane after T-duality. Since the D4-brane is a 5-brane of M-theory
wrapped on the 9th direction, we see that a non-wrapped membrane
becomes a 5-brane that is wrapped on all of the directions of $\MT{3}$ \Ah.


\subsec{T-duality in the Matrix model}

We have seen that M-theory on $\MT{3}$ is equivalent to $\SUSY{4}$
$U(N)$ SYM on the dual torus $\MHT{3}$:
\eqn\sym{\eqalign{
H =& \half\int d^3x \,{\rm tr}\{g^2 \vec{E}^2 + \inv{g^2}\vec{B}^2 
      + {\theta\over 8\pi^2}\vec{E}\cdot\vec{B} 
      + g^2 \sum_{A=1}^6 |D_t\Phi^A|^2 \cr
 &+ \inv{g^2}\sum_{A=1}^6 |D_i\Phi^A|^2
      + \inv{g^2}\sum_{1\le A<B\le 6}|\com{\Phi^A}{\Phi^B}|^2
      + ({\rm fermions})\}.\cr
}}
The coupling constant becomes
\eqn\cpl{
\tau = {4\pi i \over g^2} + {\theta \over 2\pi} = i V_\MT{3} + C
}
where $C$ is the 3-form VEV on $\MT{3}$.
The S-duality transformation 
\eqn\namee{
\tau\rightarrow -{1\over \tau}
}
agrees with \vv.
According to \tdl, the sides of $\MHT{3}$ are rescaled by 
$(\Im\tau)^{2/3}$ without changing its shape.
Thanks to the exact conformal invariance of $\SUSY{4}$ Yang-Mills
we can rescale $\MHT{3}$ and define the matrix model on a $\MHT{3}$
of volume 1. This has to be accompanied by rescaling of the six scalars
$X^\u$ by $V^{-1/3}$. The S-dual $X^\u$ should thus be rescaled 
by $V^{1/3}$. Since S-duality of SYM multiplies the Higgs VEVs
by $V=1/g^2$ we find that altogether $X^\u$ is rescaled by $V^{1/3}$.
This is in accord with the Weyl rescaling of $G_{\u\v}$ in \tdl.


\newsec{T-duality action on branes}

In this section we examine how the action of T-duality on states
is manifested in the matrix model.
Since S-duality exchanges electric and magnetic fluxes, a natural
guess is that electric (magnetic) states correspond to 
graviton (membrane) states.

As described in Section 3,
both the winding number of a membrane
and the KK momentum are parameterized by vectors in the
dual lattice with unit cell $\MHT{3}$.
Moreover, T-duality exchanges the winding and the momentum.
Electric and magnetic fluxes are also parameterized by vectors in
the dual lattice.
Their exchange under S-duality is in accord with their identification
with branes.
We shall now examine the correspondence in more detail.


\subsec{Review of electric and magnetic fluxes}

We recall that $U(1)$ is a normal subgroup of $U(N)$ and
\eqn\namef{
U(N) = (U(1)\times SU(N))/\BZ_N
}
We will consider $U(N)$ Super-Yang-Mills on $\MHT{3}$.
On $\MHT{2}\subset\MHT{3}$ we can have non-trivial $U(N)$-bundles.
States with one unit of $U(1)$ magnetic flux satisfy:
\eqn\nameg{
\int_{\MHT{2}} \trp{B} = 2\pi.
}
To build the corresponding bundle, we pick \thft:
\eqn\nameh{
U^{-1} V^{-1} U V = e^{-{2\pi i\over N}},\qquad U,V \in SU(N).
}
Let the coordinates on $\MHT{2}$ be $x,y$ with period $1$.
The bundle is defined by the boundary conditions
\eqn\bundl{
\phi(x,y) = e^{-{2\pi i y\over N}} U^{-1} \phi(x+1,y) 
 = V^{-1} \phi(x,y+1).
}

A state with one unit of electric flux in the direction of
$\MS{1}\subset \MT{3}$ (where $\MS{1}$ is some cycle of the torus)
satisfies:
\eqn\namei{
\int_{\MS{1}} \trp{E} = {2\pi}.
}


\subsec{Fluxes and compactified M(atrix)-theory}

In this section we show how
the membrane of \BFSS\ naturally becomes
a state with magnetic flux after compactification.
To see this, we begin by showing how twisted $U(N)$ bundles
are obtained from the construction \zdact.
The identification of $X^\u$ as $(i\partial^\u+ A^\u)$
at the end of Section 2 suggests that we write the same operator
as a matrix acting on sections of the {\it twisted} bundle.
Given this identification, we have 
\eqn\reltbm{
\com{X^7}{X^8} = \com{\nabla^7}{\nabla^8}
 = i B^{78}
}
where in the sector with one unit of flux, $B^{78}$ is the scalar
matrix ${2\pi\over N}\Id$.
As $N\rightarrow\infty$ it is natural to identify this state with 
the wrapped BFSS membrane.

We will now go through this construction explicitly.
We begin by finding a complete basis of sections of the twisted bundle.
Working on $\MT{2}$ and taking the sections in the fundamental
representation:
\eqn\funsec{\eqalign{
\phi(x+1,y) &= e^{{2\pi i\over N}y} U \phi(x,y),\cr
\phi(x,y+1) &= V \phi(x,y),\cr
}}
where 
\eqn\wepick{
U = 
\pmatrix{
1& & & \cr
& e^{2\pi i\over N} & & \cr
& & \ddots & \cr
& & & e^{2\pi i (N-1)\over N}
},\qquad
V =
\pmatrix{
 & 1 & & \cr
 &   & 1 & \cr
 &   &   & \ddots \cr
1&   &   &  
} }
\eqn\namej{
U^{-1}V^{-1}UV = e^{-{2\pi i\over N}}
}
We define 
\eqn\xunab{
X^\u = (\nabla^\u)^{\rm twisted} \equiv i\partial^\u +A^\u
}
The expansion of $\phi_k(x,y)$ in terms of a complete basis
is very different from the untwisted sector:
\eqn\funk{
\phi_k (x,y)
 = \sum_{p\in\BZ} \hat{\phi}(y + k + N p) e^{{2\pi i \over N}(y + k + N p)x}.
}
Here $\hat{\phi}$ is some arbitrary continuous function
defined on $(-\infty,\infty)$ with $\int|\hat{\phi}|^2 < \infty$.
So, the space on which $X^\u$ is defined is now continuous (though,
since $\hat{\phi}\in L^2(\BR)$ we can find a countable basis as before).

Writing $X^\u$ in the basis of $\hat{\phi}(w)$ we find
\eqn\xphi{\eqalign{
(X^7)_{kl}(w,w')  =&\quad {2\pi \over N} w\, \delta (w-w') \delta_{kl} \cr
 &- \sum_{p,q\in\BZ,\,s\in\BZ_N}
  \delta (w' + N p + k - l - w) 
   e^{2\pi i (q + {s\over N})w'} a^7_{p+{k-l\over N},q+{s\over N}},\cr
(X^8)_{kl}(w,w')  =& \quad i \delta'(w-w') \delta_{kl} \cr
 &- \sum_{p,q\in\BZ,\,s\in\BZ_N}
  \delta (w' + N p + k - l - w) 
   e^{2\pi i (q + {s\over N})w'} a^8_{p+{k-l\over N},q+{s\over N}},\cr
}}
Here $-\infty<w,w'<\infty$ are the arguments of $\hat{\phi}$
and $k,l$ are $U(N)$ indices.
$a^7,a^8$ are the modes of the gauge fields:
\eqn\solaa{\eqalign{
{A^7}_{kl} (x,y) &=
\sum_{p,q\in\BZ,\,s\in\BZ_N}
 {a^7}_{p+{k-l\over N},q+{s\over N}} e^{2\pi i (p + {k-l\over N})x}
     e^{2\pi i (q + {s\over N})(y+l)},\cr
{A^8}_{kl} (x,y) &=
{2\pi\over N}x \,\,+
\sum_{p,q\in\BZ,\,s\in\BZ_N}
 {a^8}_{p+{k-l\over N},q+{s\over N}} e^{2\pi i (p + {k-l\over N})x}
     e^{2\pi i (q + {s\over N})(y+l)}.\cr
}}
{}From \xunab\ it is obvious that the BFSS action goes over to
the SYM action.
The $U_1$ and $U_2$ matrices which define the sector
are still given by multiplication by
$e^{i x}$ and $e^{i y}$.  As $N\times N$ matrices they are different
from those for the untwisted sector because they act on sections \funk\
rather than functions.
Thus there are several ways to embed $\BZ^d$ subgroups which are
{\it not} mutually conjugate.

We also note that there is another way to realize the twisted
$X^\u$'s.
For compactifications on $\MT{2}$, we take a vector space
(on which $X^\u$ will act) that is a product of a finite dimensional
vector space
$V_N$ of dimension $N$ and single Hilbert space $H$ with a countably
infinite basis
\eqn\confin{
V = V_N\otimes H.}  Note the difference between \confin\ and \vvv\ --
here only {\it one} copy of $H$ (rather than two) is used for
compactification on $\MT{2}$.  This difference may be related to the
form of the expansion \funk.  We realize $U_1$ and $U_2$ as
\eqn\reuu{
U_1 = U\otimes e^{i P/\sqrt{N}},\qquad
U_2 = V\otimes e^{i Q/\sqrt{N}},\qquad
}
where $Q,P$ are canonical operators acting on $W$ 
($\com{Q}{P} = 2\pi i$) and $U,V$ are as in \wepick.
$U_1$ and $U_2$ commute and one can check that the expansion
of \zdact\ agrees with \xphi.

Next we show the relation between electric flux and graviton states.
The operator that measures the total electric flux inside $\MHT{3}$
is given by
\eqn\namek{
\int_\MHT{3}\trp{E_i}.
}
Using 
\eqn\namel{
X^\u = i\partial^\u+ A^\u,
}
we can write the electric flux as
\eqn\namem{
\int_\MHT{3}\trp{E_i} = \int_\MHT{3} \dot{A}_i
= {\bf Tr}\{\dot{X}_i\} = {\bf Tr}\{\Pi_i\}
}
where ${\bf Tr}$ is the trace in the infinite basis of the matrix model.
Thus, electric flux in SYM naturally corresponds to momentum along the 
$i$-th direction of the $\MT{3}$ in M-theory.


\newsec{More on membranes}

In this section we describe further evidence in favor of the
identifications between magnetic flux and Lorentzian membranes.
We also propose a connection between Yang-Mills instantons and
Euclidean membranes.


\subsec{Energy and counting of wrapped membrane states. } 

Consider M-theory on $T^2$.
We would like to calculate the energy of the membrane 
using the fact that the membrane states correspond to 
states in the sector with non-zero magnetic flux
in the $U(1)$ subgroup. 
In ``mathematical'' conventions where there is an overall 
$g^{-2}_{YM}$ in front of the action, the unit 
of flux is 
\eqn\unfl{ \int tr B = 2\pi.} 
Taking $B$ of the form $B_0$  times the identity 
matrix, we find that 
\eqn\namen{
B_0 = {1 \over {2\pi N  R_1^{YM}R_2^{YM}}} } 
where T-duality relates the Yang-Mills and M-theory lengths through
$R_i^{YM} = {2 \pi \alpha^{\prime} \over R_i}$.
In this subsection we restore the 11D Planck length

\eqn\nameo{ 
l_p = {g}^{1/3} \sqrt{\a'}.
}
The coupling constant of the YM   is given by
\eqn\ymcou{  g_{YM}^{2}= 
{g\sqrt{\alpha'}  \over {R_1R_2} }.}
Calculating the energy we get : 
\eqn\energ{{1\over 2} g_{YM}^{-2} \int tr B^2 = { (2\pi)^{-4}
 (R_1R_2)^2 \over {
2gN (\alpha^{\prime} )^{5/2} }} }

 As in the  discussion of the membrane tension in \BFSS, 
the energy of the membrane in the light-cone frame  
is   given by 
\eqn\enlc{ E = \sqrt { p^2 + M^2} = 
\sqrt { \bigl( {N\over  g\sqrt { \alpha^{\prime}} } \bigr )^2 
            + (TR_1R_2)^2} 
             \sim { N\over {g\sqrt{ \alpha^{\prime}}} }
                + {g \sqrt{ \alpha^{\prime} } T^2 (R_1R_2)^2\over 2 N}.}
Here $T$ is the membrane tension and 
$ { N \over {g\sqrt {\alpha^{\prime} } }} $
is the momentum along $X_{11}$.
 The energy of the state with magnetic 
flux in the large $N$ $U(N)$ Yang Mills
is interpreted as the excess kinetic energy.
The total energy is obtained by adding the term
$N/g\sqrt{\a'}$ due to the momentum along $R_{11}$
to the kinetic energy.
This explains why the
square of the area appears in the numerator, and  gives the 
tension of the membrane, 
\eqn\memten{ T^2 = {1\over { (2\pi)^4 g^2 \alpha^{\prime}}^3} =  
{1\over {(2\pi)^4 l_p^6 } }}  
in agreement with \BFSS. 

M-theory on $R^9 \times T^2$ should have, for fixed momentum on $R^9$,
one normalizable BPS multiplet (annihilated by half the
supersymmetries) with the quantum numbers of an $n$-wrapped membrane.
This is known \sch\ by using the relation between M-theory and type II
strings.  In the context of the relation between M-theory and large
$N$ $U(N)$ Yang Mills, and the above description of membrane winding
number as the magnetic flux in the $U(1)$ we can argue that this is
the right counting.  So we need to know the number of states coming
from large N $U(N)$ Yang Mills which sit in a $2^8$ dimensional
representation of supersymmetry and carry $n$ units of magnetic flux.
Following arguments in \WitBST\ and in \town, and in agreement with
the discussion at the end of Section 4, precisely this Yang Mills
question arises if we want to count the number of normalizable bound
states carrying $N$ units of 2-brane charge and $n$ units of 0-brane
charge in type IIA theory. We know that the unique bound state
\senmarg\ of $N$ type IIA 2-branes can be given $n$ units of momentum
along the eleventh dimension. We conclude that there should be a
unique $1/2$ BPS saturated state coming from the sector with $n$ units
of magnetic flux in $U(N)$ Yang Mills. This statement in the large N
limit shows that M-theory on $T^2$ has exactly one BPS multiplet
carrying the charge of a membrane wrapped $n$ times on the $T^2$.

\subsec{Euclidean membranes}
Since Lorentz invariance is not manifest in the BFSS formulation,
it is interesting to check what an instantonic membrane looks like.
Since the Minkowski metric is crucial to the IMF formulation,
we will think of a Euclidean membrane as a transition between two
different vacua which we will soon identify.
Let's take a Euclidean membrane that wraps all of $\MT{3}$.
Since it is localized in directions $1\dots 6$, the corresponding SYM
solution must have all six adjoint scalars set to a scalar matrix, say:
\eqn\namep{
\Phi^1 = \cdots = \Phi^6 = 0.
}
This allows for an ordinary Yang-Mills instanton solution 
in the gauge fields.
It is thus tempting to guess that a fully wrapped membrane corresponds
to a transition between two vacua which differ by a large gauge
transformation related to $\pi_3(U(N))$.
In favor of this claim we recall from \vv\ that the $\theta$-angle
is related to $C_{7,8,9}$. A membrane instanton that wraps $\MT{3}$
is a transition in which (in an appropriate gauge):
\eqn\nameq{
\int dx_{1-6} dx_{11} {d\over dt}C_{7,8,9} \sim {d\over dt}\theta
}
changes by one unit. If we replace ${d\over dt}\theta$ by
the dual variable we find that in the SYM formulation
the winding number of the vacuum increases by one unit.
The action of the YM instanton, ${4\pi\over g^2}+{i\theta\over 2\pi}$,
agrees with
the action $V_{7,8,9}+ i C_{7,8,9}$ of the membrane instanton.

\newsec{5-branes in M(atrix) theory}

We now discuss several issues related to the construction of a 5-brane
in M(atrix) theory.  A 5-brane which extends along the light-cone
directions $x^\pm$ (and four more directions) has been discussed in
\BerDou.  However, the 5-brane described by these authors was
essentially given as a background for the 0-brane theory.  By using
the relation between the 0-brane fields on a torus and the covariant
derivative operator on the dual torus, we find a natural description
of the light-cone 5-brane which is intrinsic to the 0-brane variables
of M(atrix) theory.  We also discuss the possibility of using
T-duality to describe a 5-brane that occupies five transverse
directions and is boosted along the BFSS preferred direction.

\subsec{The light-cone 5-brane}

It was shown by Witten \witteni, and in a more general form by 
Douglas \douglas, that an instanton 
on a $(p + 4)$-brane carries $p$-brane charge.
This result is essentially due to the fact that the world volume
theory on the $(p + 4)$-brane includes a Chern-Simons term
\eqn\namer{
\int_{\Sigma_{p+ 5} } C\wdg e^F.
}
where $C$ is a sum over RR fields.
The term
\eqn\names{
\int C^{(p + 1)}\wdg F \wedge F
}
in particular couples an instanton to the field $C^{(p + 1)}$, under
which $p$-branes are electrically charged.  As one application of this
result, Yang Mills instantons embedded in a three-brane world-volume
theory appear as D-instantons. After a T-duality which converts three
branes to zero branes, these D-instantons are converted to membrane
instantons.  Relating this to the transformation from the 0-brane
variables of M-theory on  $T^3$ to the Yang Mills variables, this
gives another piece of evidence
in favour of the identification proposed in section 5.2.

We can use this same type of relation to show how a 5-brane of
M-theory can be constructed directly from the $X$ matrix fields of
BFSS.  Wrapping the 5-brane around the light-cone directions, we
should find a 4-brane in the resulting IIA theory.  Let us compactify
in dimensions 6-9.
Since a 4-brane
on a torus $\MT{4}$ goes to a 0-brane on the dual torus  $\MHT{4}$,
the connection \nabdef\ indicates that a configuration of the $X$
matrices satisfying
\eqn\five{
{\bf Tr}\; \epsilon_{ijkl} X^i X^j X^k X^l=8 \pi
}
will have a unit of 4-brane charge.  (The indices $i-l$ are summed
over the compactified coordinates 6-9.
The antisymmetric product of 4
$X$'s is just ${1\over 4}\epsilon_{ijkl}F^{ij}F^{kl}$ on the dual torus
written in 0-brane language.)
This condition for a 5-brane configuration is very similar to the
condition
\eqn\namet{
{\bf Tr}\; \epsilon_{ij} X^i X^j = 2 \pi i
}
for a membrane,  which as in \reltbm\ is essentially the description
given in \refs{\BFSS,\NDW}.  Just as for the membrane, it is natural to
expect that in the decompactified limit, \five\ will be satisfied for
a light-cone 5-brane.  As for the 2-brane, this condition cannot be
satisfied by finite dimensional matrices.

It is interesting to note that the trace in \five\ is nonvanishing
when there are membranes wrapped around for example the 67 and 89
directions.  However, in this case the trace scales as $1/N$ and
vanishes in the large $N$ limit.  An explicit example of a
configuration where
\five\ is satisfied can be constructed by simply taking the matrices
corresponding to the action of the operators $i \partial^\mu + A^\mu$,
where $A^\mu$ are gauge fields corresponding to an instanton on
$\MHT{4}$.  Such a configuration presumably corresponds to a single
5-brane in M(atrix) theory.

\subsec{Transverse 5-branes}

Since T-duality transforms a 5-brane that wraps $\MT{3}$ into a membrane
that is unwrapped we can relate the wave-function of a wrapped 5-brane
to a wave-function for an unwrapped 2-brane through
\eqn\wrfiv{
\ket{{\rm 5-brane},\, i,j,7,8,9} = S\ket{{\rm membrane},\, i, j}
}
where $S$ is the S-duality operator and
$\ket{{\rm membrane},\, i, j}$ is a state of $U(\infty)$ SYM in which 
the scalars $\Phi^i,\Phi^j$ (two of the six adjoint scalars) have
condensed in the form
\eqn\phicond{\eqalign{
\Phi^i &= P + ({\rm oscillators}),\cr
\Phi^j &= Q + ({\rm oscillators}),\cr
}}
where $P,Q$ are canonical $\infty\times\infty$ matrices.

The usefulness of formula \wrfiv, of course, depends upon
progress in finding an explicit form for the 
S-duality operator $S$ and its generalization to $U(\infty)$.
Once we have the wave function of the 5-brane on $\MT{3}$ we can
decompactify to 11D M-theory.

{}From \cpl\ we see that this is done by taking the coupling constant
of SYM to zero. At first sight one might worry that this gives a classical
theory. However, we must first take $N\rightarrow\infty$ and only then
take $g\rightarrow 0$. Thus, the effective coupling constant $g^2 N$
is never perturbative.

\newsec{Conclusions}

We have seen that T-duality is realized in a natural way in the
M(atrix)-theory of BFSS as S-duality of large $N$
$U(N)$ $\SUSY{4}$ Yang-Mills theory.
The large/small volume duality of M-theory is mapped
to the weak/strong coupling duality of SYM
and graviton~(0-brane)/membrane duality is mapped to
electric/magnetic duality.
We have also seen that there are different inequivalent embeddings
of the $\BZ^d$ symmetry group of translations in the BFSS model.
The different embeddings give rise to different $U(N)$ bundles
in the SYM theory. The wrapped membrane of M-theory is identified
with states carrying magnetic flux.

The connections developed in this paper provide a natural framework in
which to try to understand the 5-brane in M(atrix) theory.  The
unwrapped 5-brane is naturally related through S-duality of 3+1
dimensional Super-Yang-Mills theory to a 2-brane configuration which
can be understood in matrix variables $X$.  Furthermore, the 5-brane
wrapped around the light-cone directions has a natural description as
an ``instanton'' of 4+1 dimensional Super-Yang-Mills, which allows
us to describe it in terms of a set of matrix variables satisfying the
relation ${\bf Tr}\; \epsilon_{ijkl} X^i X^j X^k X^l= 8 \pi$.  In the
language of type IIA string theory, the fact that it is possible to
describe the
4-brane in terms of fundamental 0-brane fields is 
essentially the T-dual of the result that instantons on a 4-brane
carry 0-brane charge.  Remarks along these lines were also made in \sg.
It would seem that the ability of 0-branes to
form the higher dimensional branes of M-theory is a strong argument in
favor of the conjecture of BFSS that in fact 0-branes form a complete
description of all the degrees of freedom in M-theory, at least in the
IMF frame.

\bigbreak\bigskip\bigskip
\centerline{\bf Acknowledgments}\nobreak
We wish to thank A. Hashimoto, R. Gopakumar, D. J. Gross, I. Klebanov,
G. Lifschytz, S. Mathur and E. Witten for helpful discussions.  OJG is
also grateful to T. Banks for a discussion during the Rutgers
University theory group meeting.  The research of SR and WT is
supported by NSF Grant PHY96-00258 and the research of OJG is
supported by a Robert H. Dicke fellowship and by DOE grant
DE-FG02-91ER40671.

\bigbreak\bigskip\bigskip
\centerline{\bf Note added}\nobreak
As this work was completed, a paper by L. Susskind \Sus\ appeared which
discusses the same issue.

\listrefs

\end